\documentclass{iaus}
\usepackage{graphicx,epsfig}

\title[Building Disks in Triaxial Halos] %% give here short title %%
{Building Galactic Disks\\ in Triaxial Dark Matter Halos}

\author[Isaac Shlosman] %% give here short author list %%
{Isaac Shlosman}

\affiliation{Department of Physics \& Astronomy, University of Kentucky,
Lexington, KY 40506, USA \break email: shlosman@pa.uky.edu}

\pubyear{2006}
\volume{235}  %% insert here IAU Symposium No.
\pagerange{1--4}
\date{?? and in revised form ??}
\jname{Galaxy Evolution across the Hubble Time}
\editors{F.Combes \& J. Palous, eds.}
\begin{document}

\maketitle

\begin{abstract}
We review our recent work on the formation and evolution of disks within triaxial 
dark matter (DM) halos by means of numerical simulations, including star formation 
and feedback from stellar evolution. The growing disks are strongly influenced by
shapes of DM halos and modify them in turn. Disk parameters are in a broad agreement
with those in the local universe. Gas-rich stellar bars grow in
tandem with the disk and facilitate the angular momentum redistribution in the system and
radial gas inflow. Nested bars appear to form as a by-product. Interactions between 
various non-axisymmetric components --- bars, disks and halos lead to decay of bars or 
washing out of ellipticity in the inner halo.
%% add here a maximum of 10 keywords, to be taken from the file <Keywords.txt>
\end{abstract}

\firstsection % if your document starts with a section,
\section{Introduction}

% NOTE use of \upi in above paragraph and subsequently throughout paper.
% The Greek constant character pi should be upright.
 
Fundamental difficulties remain in our understanding of structure formation on the
subgalactic scales. Among these, formation and evolution of galactic disks and bulges, 
angular momentum redistribution between baryonic and dark matter (DM), radial density 
profiles of DM halos (especially survival of the DM cusps), fueling of the
central activity, correlation between masses of the central black holes (SBHs)
and the surrounding host galaxies --- provide an incomplete but impressive
list of stand-by problems. On the other hand, understanding the origin of the 
large-scale structure in the universe has been largely successful. 

Because the formation and evolution of galactic disks involves grossly non-linear 
processes, both dissipative and non-dissipative, they are typically investigated by means 
of extensive numerical simulations. A major outcome of these simulations is that halos 
forming within the $\Lambda$CDM cosmology appear to be universally triaxial (e.g., 
Allgood et al. 2006) while those inferred in the nearby universe are compatible 
with being axisymmetric (e.g., Combes 2002). Here we argue that the underlying halo 
shape evolution is directly related to the disk formation and its subsequent growth. 

Two main approaches have been used to follow the formation and evolution of disk
galaxies. First, large-scale cosmological simulations of collisionless and 
collisional matter have been invoked, with a subsequent extraction of a DM halo and 
its immediate environment. Resampling was implemented in order to increase the 
number of particles per halo (e.g., Sommer-Larsen et al. 2003; Governato et al. 2006). 
With all the advantages, these simulations still suffer from insufficient numerical 
resolution and cannot compete with simulations of individual galaxies. 
Second, models of galactic disks in frozen spherically-symmetric halos or 
analytically growing (frozen) halos have been advanced (e.g.,
Samland \& Gerhard 2003; Immeli et al. 2004). They treat a variety of
important processes, such as chemical evolution, but cannot follow other processes,
such as angular momentum redistribution in the system, especially the halo-disk resonant
coupling. 

Here we summarize our results from two alternative approaches: (1) simulations of a 
limited volume using cosmological initial conditions obtained by means of Constrained 
Realizations method (Hoffman \& Ribak 1991) of a prescribed density field (Romano-Diaz 
et al. 2006, 2007). Berentzen \& Shlosman (2006) have applied this method in order to 
follow the 
evolution of DM within a $4h^{-1}$~Mpc$^3$ box, starting with a redshift $z=120$ and 
inserting a `seed' disk at $z=3$, in the open CDM (OCDM) Universe with $\Omega_0=0.3,
h=0.7$ and $\sigma_8=0.9$. The disk was grown tenfold over different periods of 
time, $1-3$~Gyr. (2) Heller, Shlosman \& Athanassoula
(in preparation) have followed the Hubble expansion and collapse of an isolated
DM$+$baryons perturbation starting with $z=36$. Our star formation (SF) algorithm is 
physically motivated and feedback
from stellar evolution in the form of OB stellar winds and SN is included. We have
used the updated hybrid SPH/$N$-body code FTM-4.4 of Heller \& Shlosman.  

\section{Galaxy Models from Constrained Realizations}

The DM model has been constructed with a nested set of three off-center $2.5-3.5\sigma$ perturbations. The pure DM halo (model A0) went through the epoch of major mergers 
which ended at $z\sim 7$.
Its density profile is given by NFW (Navarro et al. 1997) established by $z\sim 10$.
We use the {\it isopotential} contours to measure the halo shape, which has a substantial
advantage over the isodensities method (Berentzen et al. 2006). The A0 model has settled
in a stable, substantially triaxial (elliptical and flat) configuration (Fig.~1). 
This halo ellipticity leads to a strong interaction with a seed disk
immersed at $z=3$ and grown (or not) over the next 1--3~Gyr.  

\begin{figure}
 \begin{center}
  \includegraphics[angle=-90,scale=0.6]{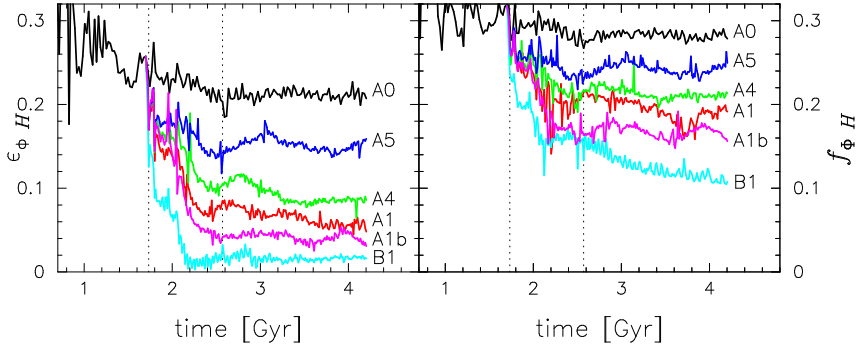}
   \end{center}
    \caption{Evolution of the halo equatorial {\it potential} ellipticity (left) and 
{\it potential} flatness(right) at representative 5~kpc in different model realizations: 
A0 (pure DM), A1 (growing disk), A4 (slowly growing disk), A5 (nongrowing seed disk), A1b 
(frozen disk), B1 (massive growing disk). Vertical dotted lines (from left to right): $z=3$ 
and 2.} 
\end{figure}
 
We find that a growing disk is responsible for washing out the halo equatorial ellipticity 
(in the disk plane) and for diluting its flatness over a period of time comparable to the 
disk growth. In models where the disk contributes more to 
the overall rotation curve, it appears more efficient in axisymmetrizing the 
halo, {\it without} accelerating the halo figure rotation.
The observational corollary is that 
the maximal disks probably reside in nearly axisymmetric halos, while disks whose 
rotation is dominated by the halo at all radii can reside in more elliptical halos. 
The halo shape is sensitive to the final disk mass, but is independent of how the seed 
disk is introduced into the system --- abruptly or quasi-adiabatically. It is only weakly 
sensitive to the timescale of the disk growth. While models with massive disks have
developed strong bars, we find that light disks have this instability damped by the halo 
triaxiality --- an interesting implication for the cosmological evolution of disk
galaxies. The halo itself responds to the stellar bar by developing a gravitational wake 
--- a `ghost' bar of its own which is almost in-phase with that in the disk.

\section{Galaxy Models from Isolated Perturbations} 
 
We have also investigated formation and evolution of galactic disks
within assembling live DM halos in a large number of models (Heller et al., in 
preparation). Disk and halo 
structural components have been evolved from the cosmological initial conditions and 
represent the collapse of an isolated density perturbation in the OCDM Universe with 
$\Omega_0=0.3$, $h=0.7$ and 10\% baryons. The gas temperature was obtained from the energy equation. The SF takes place in the Jeans unstable, contracting
region. We fix the gas-to-background density at which the gas is converted
to a star, the local collapse-to-free fall time, and introduce the
probability that the gas particle produces a star during a given timestep.
Four generations of stars form per gas particle with an instantaneous    
recycling along with an increment in metallicity. 
The balance of the specific internal energy along with the gas 
ionization fractions of H and He and the mean molecular weight are 
computed as a function of $\rho$ and $T$ for an optically thin 
primordial composition gas.

\begin{figure}
 \begin{center}
 \psfig{figure=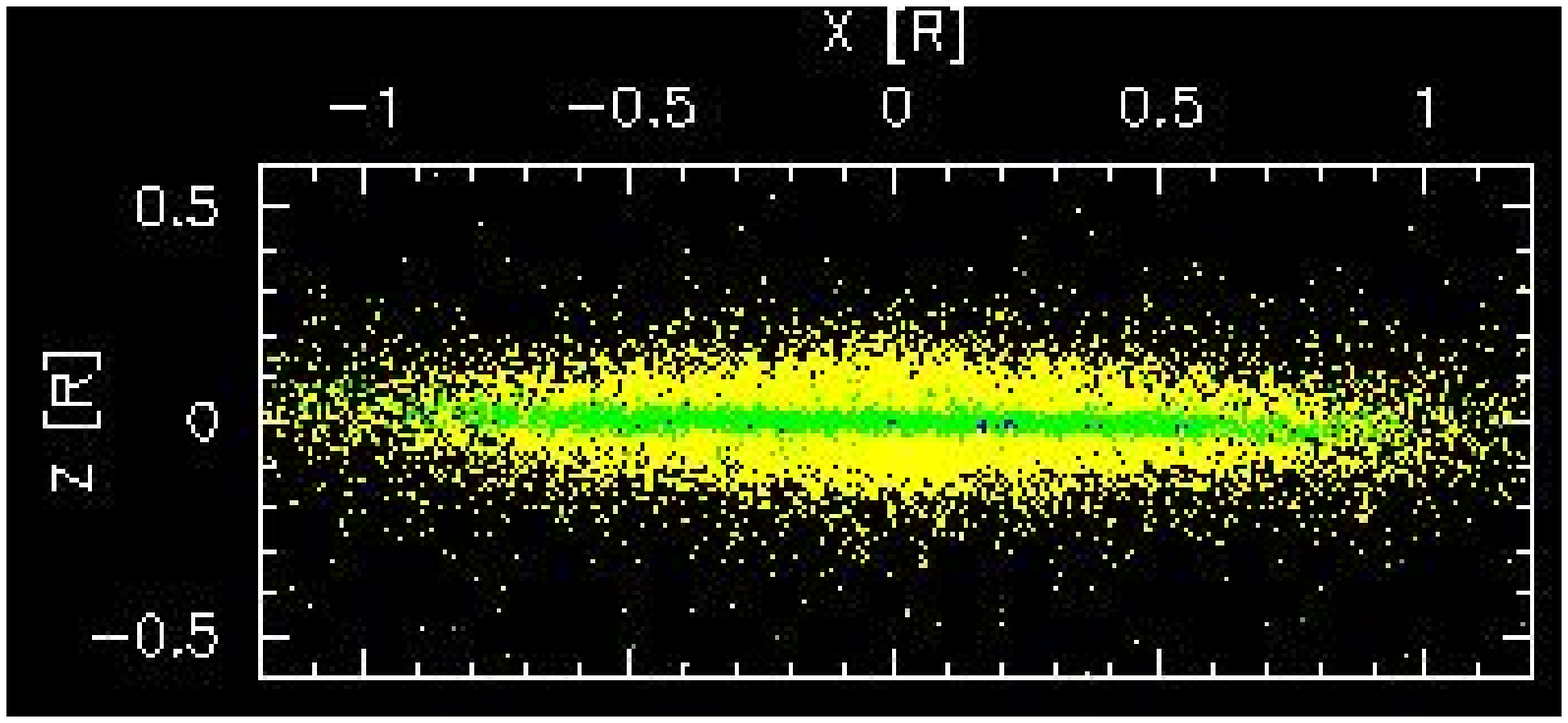,width=6.62cm,angle=0}\hspace{0.001cm}
\psfig{figure=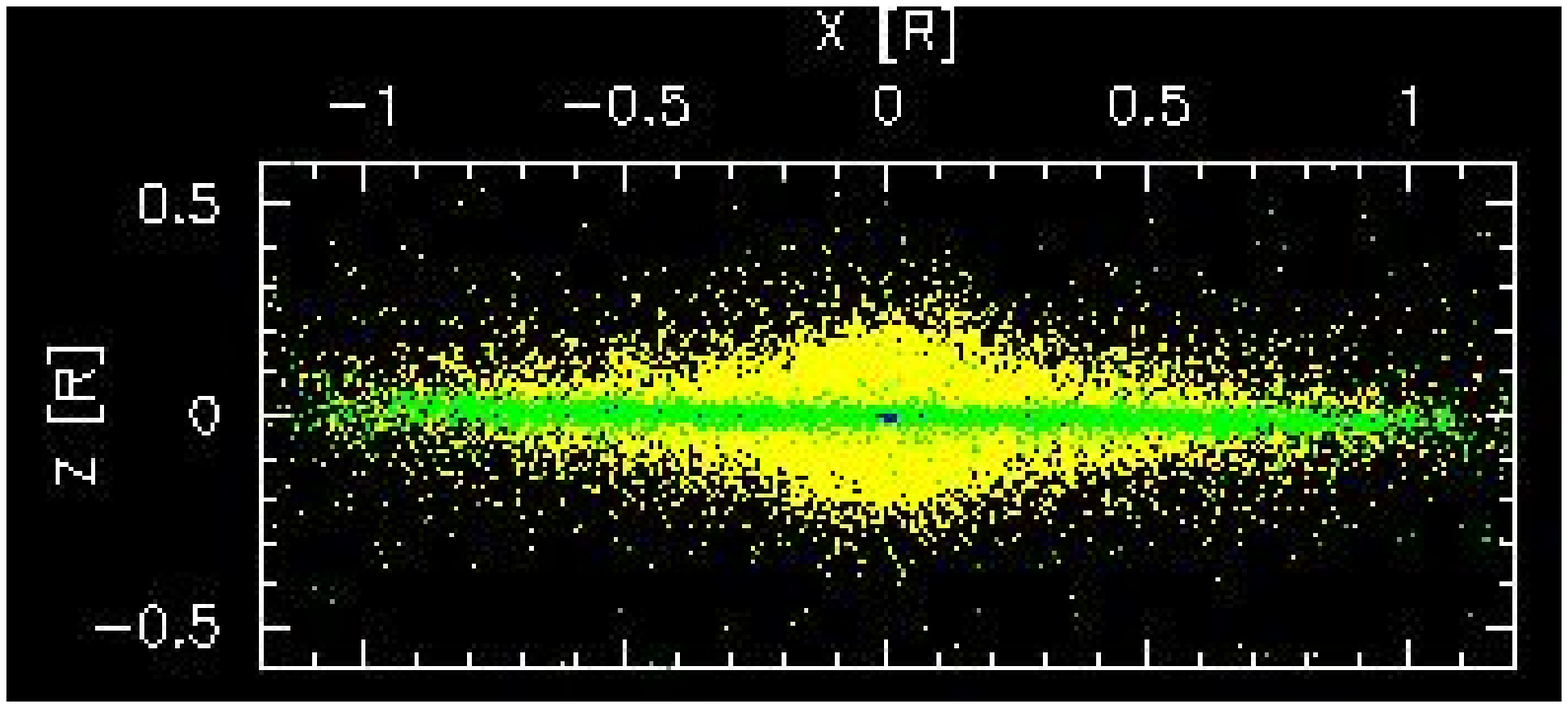,width=6.75cm,angle=0}\\
   \end{center}
    \caption{Examples of nearly bulgeless (left) and bulge-dominated (right) edge-on disks.
    Colors: green (gas), yellow (stars), blue (star formation), DM (omitted). Units: [1] =
    10~kpc.}
\end{figure}

The energy feedback from the stellar evolution includes the OB stellar winds and SN, and 
uses the thermalization parameter
--- a fraction of the feedback deposited as a thermal energy and converted
into kinetic energy via equations of motion. Full details of the SF algorithm and
energy feedback are given elsewhere. The initial density profile corresponds to the 
average density 
around a $2\sigma$ peak in a Gaussian random density field. The spin parameter 
is set to $\lambda = 0.05$. 
We impose the collapse redshift of $z=2$. The mass of collisionless DM particles is 
$7\times 10^{11}~{\rm M_\odot}$ and the gas comprises 10\% of the total mass initially.

We find that the pure DM models lead to the formation of triaxial halos with a negligible
tumbling of $\sim \pi$ over the Hubble time. This leads to corotation and
inner Lindblad resonance being moved to very large and small radii,
respectively. Orbital response of a disk immersed in such a halo will be always out
of phase with the halo major axis --- thus diluting the halo equatorial potential
and washing out its ellipticity. (Baryon effect on the halo shape was first noticed by 
Dubinski [1994] by diluting the halo potential with that of a rigid Plummer sphere, and 
by Kazandzidis et al. [2004] by damping baryons inside the halo.) In addition, the 
increasing central mass concentration in the model contributes to the DM orbit 
scattering and so to decrease in halo ellipticity and flattening.

Evolution of the DM halos differs in the presence of baryons in that it reaches
smaller triaxialities and these are reduced further. Consequently, the growing
disks exhibit substantial ellipticities in the first 2~Gyr. These heavily gas rich 
disks respond dramatically to halo's non-axisymmetric driving by forming a pair
of strong grand-design density waves in the stellar component and associated shocks in 
the gas. The pattern speeds of the oval disks and the barlike response are identical
and lead to a rapid gas inflow toward the central kpc. In a number of models, this
resulted in decoupling the central part of the bar in the prograde direction --- i.e.,
formation of nested bars (Heller, Shlosman \& Athanassoula 2006) in accord with
Shlosman et al. (1989, 1990) but under different initial conditions (Begelman et al.
2006). 

The disk grows from inside out by increasing its radial scalelength. This and other disk 
parameters appear to lie within the range observed in the local universe --- from 
essentially bulgeless to bulge dominated disks (Fig.~2). They do not `suffer' from
the angular momentum `catastrophe.'  The total $J$ is approximately conserved, even in 
models accounting for feedback from the stellar evolution. The baryons lose $\sim 
30\%$ and the DM acquires $\sim 3\%$ of their original $J$. The specific
$J$ for the DM is nearly constant while that for the baryons is decreasing with time. 

\underline{In summary,} shapes of host DM halos are shown to have a profound effect on 
the morphology and 
dynamics of growing disks --- with the caveat that the disk growth alters the halo shape. 
We find that maximal disks wash out the equatorial ellipticity of inner ($< 20$~kpc) halos
and lessen their flatness. They are expected to reside in (nearly) axisymmetric halos
in the local universe, while disks dominated by halos at all radii can reside 
in elliptical halos. The initial disk response to the halo asymmetry is barlike. This
facilitates the angular momentum transfer between the disk and the host halo, and
triggers the radial gas inflow. These conditions provide fertile grounds for the formation
of gas rich nested bars. Whether the large-scale bars survive for longer time periods 
depends largely on the timescale of a halo losing its ellipticity, as stellar 
bars and elliptical halos appear to be incompatible (El-Zant \& Shlosman 2002).

\begin{acknowledgments}
I am indebted to my collaborators Lia Athanassoula, Ingo Berentzen, Clayton Heller,
Yehuda Hoffman, Shardha Jogee and Emilio Romano-Diaz on these issues. The relevant grants from NASA and NSF are 
gratefully acknowledged.
\end{acknowledgments}

%\begin{discussion}

%\end{discussion}

\begin{thebibliography}{}

\bibitem[]{}Allgood, B., et al. 2006, MNRAS, 367, 1781

\bibitem[]{}Begelman, M.C., Volonteri, M. \& Rees, M.J. 2006, MNRAS, 370, 289

\bibitem[]{}Berentzen, I., Shlosman, I. \& Jogee, S. 2006, ApJ, 637, 582

\bibitem[]{}Berentzen, I. \& Shlosman, I. 2006, ApJ, 648, 807

\bibitem[]{}Combes, F. 2002, New Astron. Rev., 46, 755

\bibitem[]{}Dubinski, J. 1994, ApJ, 431, 617

\bibitem[]{}El-Zant, A. \& Shlosman, I. 2002, ApJ, 577, 626

\bibitem[]{}Governato, F., et al. 2006, MNRAS, submitted (astro-ph/0602351)

\bibitem[]{}Heller, C.H., Shlosman, I. \& Athanassoula, E. 2006, ApJ Lett., subm. (astro-ph/0610428)
 
\bibitem[]{}Hoffman, Y. \& Ribak, E. 1991, ApJ, 380, L5

\bibitem[]{}Immeli, A., et al. 2004, A\&A, 413, 547
     
\bibitem[]{}Kazandzidis, S., et al.  2004, ApJ, 611, L73

\bibitem[]{}Navarro, J.F., Frenk, C.S. \& White, S.D.M. 1997, ApJ, 490, 493

\bibitem[]{}Romano-Diaz, E., et al. 2006, ApJ, 637, L93  

\bibitem[]{}Romano-Diaz, E., et al. 2007, ApJ, 655, in press (astroph/0610090)

\bibitem[]{}Samland, M. \& Gerhard, O. 2003, A\&A, 399, 961   
     
\bibitem[]{}Shlosman, I., Frank, J. \& Begelman, M.C. 1989, Nature 338, 45

\bibitem[]{}Shlosman, I., Begelman, M.C. \& Frank, J. 1990, Nature 345, 679

\bibitem[]{}Sommer-Larsen, J., Gotz, M. \& Portinari, L. 2003, ApJ, 596, 47

 

\end{thebibliography}
\end{document}